# Abstracting Definitional Interpreters

Functional Pearl

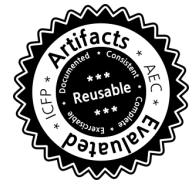

DAVID DARAIS, University of Maryland
NICHOLAS LABICH, University of Maryland
PHÚC C. NGUYỄN, University of Maryland
DAVID VAN HORN, University of Maryland

In this functional pearl, we examine the use of definitional interpreters as a basis for abstract interpretation of higher-order programming languages. As it turns out, definitional interpreters, especially those written in monadic style, can provide a nice basis for a wide variety of collecting semantics, abstract interpretations, symbolic executions, and their intermixings.

But the real insight of this story is a replaying of an insight from Reynold's landmark paper, *Definitional Interpreters for Higher-Order Programming Languages*, in which he observes definitional interpreters enable the defined-language to inherit properties of the defining-language. We show the same holds true for definitional *abstract* interpreters. Remarkably, we observe that abstract definitional interpreters can inherit the so-called "pushdown control flow" property, wherein function calls and returns are precisely matched in the abstract semantics, simply by virtue of the function call mechanism of the defining-language.

The first approaches to achieve this property for higher-order languages appeared within the last ten years, and have since been the subject of many papers. These approaches start from a state-machine semantics and uniformly involve significant technical engineering to recover the precision of pushdown control flow. In contrast, starting from a definitional interpreter, the pushdown control flow property is inherent in the meta-language and requires no further technical mechanism to achieve.

## 1 INTRODUCTION

An abstract interpreter is intended to soundly and effectively compute an over-approximation to its concrete counterpart. For higher-order languages, these concrete interpreters tend to be formulated as state-machines (e.g. Jagannathan and Weeks (1995); Jagannathan et al. (1998); Wright and Jagannathan (1998); Might and Shivers (2006a); Midtgaard and Jensen (2008); Midtgaard and Jensen (2009); Might and Van Horn (2011); and Sergey et al. (2013)). There are several reasons for this choice: they operate with simple transfer functions defined over similarly simple data structures, they make explicit all aspects of the state of a computation, and computing fixed-points in the set of reachable states is straightforward. The essence of the state-machine based approach was distilled by Van Horn and Might in their "abstracting abstract machines" (AAM) technique, which provides a systematic method for constructing abstract interpreters from standard abstract machines like the CEK- or Krivine-machines (Van Horn and Might 2010). Language designers who would like to build abstract interpreters and program analysis tools for their language can now, in principle at least, first build a state-machine interpreter and then turn the crank to construct the approximating abstract counterpart.

A natural question to arise from this past work is to wonder: can a systematic abstraction technique similar to AAM be carried out for interpreters written, *not* as state-machines, but instead as high-level definitional interpreters, i.e. recursive, compositional evaluators? This functional pearl answers in the affirmative and demonstrates some of the interesting consequences of doing so.

First, we show the AAM recipe can be applied to definitional interpreters with only a slight adaptation of the original method. The primary technical challenge in this new setting is handling interpreter fixed-points in a way that is both sound and always terminates—a naive abstraction of fixed-points will be sound but isn't always terminating, and a naive use of caching for fixed-points will guarantee termination but is inherently unsound. We address this technical challenge with a caching fixed-point-finding algorithm which is both sound and guaranteed to terminate when abstracting arbitrary definitional interpreters.

Second, we claim that the abstract definitional interpreter perspective is fruitful in two regards. The first is unsurprising: high-level abstract interpreters offer the usual beneficial properties of their concrete counterparts in terms of being re-usable and extensible. In particular, we show that abstract interpreters can be structured with monad transformers to good effect. The second regard is more surprising, and we consider its observation to be the main contribution of this pearl.

Definitional interpreters, in contrast to abstract machines, can leave aspects of computation implicit, relying on the semantics of the defin*ing*-language to define the semantics of the defin*ed*-language, an observation made by Reynolds (1972) in his landmark paper, *Definitional Interpreters for Higher-order Programming Languages*. For example, Reynolds showed it is possible to write a definitional interpreter such that it defines a call-by-value language when the metalanguage is call-by-value, and defines a call-by-name language when the metalanguage is call-by-name. Inspired by Reynolds, we show that *abstract* definitional interpreters can likewise inherit properties of the metalanguage. In particular we construct an abstract definitional interpreter where there is no explicit representation of continuations or a call stack. Instead the interpreter is written in a straightforward recursive style, and the call stack is implicitly handled by the metalanguage. What emerges from this construction is a total abstract evaluation function that soundly approximates all possible concrete executions of a given program. But remarkably, since the abstract evaluator relies on the metalanguage to manage the call stack implicitly, it is easy to observe that it introduces no approximation in the matching of calls and returns, and therefore implements a "pushdown" analysis (Earl et al. 2010; Vardoulakis and Shivers 2011), all without the need for any explicit machinery to do so.

**Outline**

In the remainder of this pearl, we present an adaptation of the AAM method to the setting of recursively-defined, compositional evaluation functions, a.k.a. definitional interpreters. We first briefly review the basic ingredients in the AAM recipe (section 2) and then define our definitional interpreter (section 3). The interpreter is largely standard, but is written in a monadic and extensible style, so as to be re-usable for various forms of semantics we examine. The AAM technique applies in a basically straightforward way by store-allocating bindings and soundly finitizing the heap. But when naively run, the interpreter will not always terminate. To solve this problem we introduce a caching strategy and a simple fixed-point computation to ensure the interpreter terminates (section 4). It is at this point that we observe the interpreter we have built enjoys the "pushdown" property *à la* Reynolds—it is inherited from the defining language of our interpreter and requires no explicit mechanism (section 5).

Having established the main results, we then explore some variations in brief vignettes that showcase the flexibility of our definitional abstract interpreter approach. First we consider the widely used technique of so-called "store-widening," which trades precision for efficiency by modelling the abstract store globally instead of locally (section 6). Thanks to our monadic formulation of the interpreter, this is achieved by a simple re-ordering of the monad transformer stack. We also explore some alternative abstractions, showing that due to the extensible construction, it's easy to experiment with alternative components for the abstract interpreter. In particular, we define an alternative interpretation of the primitive operations that remains completely precise until forced by joins in the store to introduce approximation (section 7). As another variation, we explore computing a form of symbolic execution as yet another instance of our interpreter (section 8). Lastly, we show how to incorporate so-called "abstract garbage collection," a well-known technique for improving the precision of abstract interpretation by clearing out unreachable store locations, thus avoiding future joins which cause imprecision (section 9). This last variation is significant because it demonstrates that even though we have no explicit representation of the stack, it is possible to compute analyses that typically require such explicit representations in order to calculate root sets for garbage collection.

Finally, we place our work in the context of the prior literature on higher-order abstract interpretation (section 10) and draw some conclusions (section 11).

**Style**

To convey the ideas of this paper as concretely as possible, we present code implementing our definitional abstract interpreter and all its variations. As a metalanguage, we use an applicative subset of Racket (Flatt and PLT 2010), a dialect of Scheme. This choice is largely immaterial: any functional language would do. However, to aide extensibility, we use Racket's *unit* system (Flatt and Felleisen 1998) to write program components that can be linked together.

All of the code presented in this paper runs; this document is a literate Racket program. We have also implemented a small DSL for composing and experimenting with these interpreters easily. Source code, documentation, and a brief tutorial are available at the following URL:

https://github.com/plum-umd/abstracting-definitional-interpreters

## 2 FROM MACHINES TO COMPOSITIONAL EVALUATORS

In recent years, there has been considerable effort in the systematic construction of abstract interpreters for higher-order languages using abstract machines—first-order transition systems—as a semantic basis. The so-called *Abstracting Abstract Machines* (AAM) approach to abstract interpretation (Van Horn and Might 2010) is a recipe for transforming a machine semantics into an easily abstractable form. The transformation includes the following ingredients:

- Allocating continuations in the store;
- Allocating variable bindings in the store;
- Using a store that maps addresses to *sets* of values;
- Interpreting store updates as a join; and
- Interpreting store dereference as a non-deterministic choice.

These transformations are semantics-preserving due to the original and derived machines operating in a lock-step correspondence. After transforming the semantics in this way, a *computable* abstract interpreter is achieved by:

- Bounding store allocation to a finite set of addresses; and
- Widening base values to some abstract domain.

After performing these transformations, the soundness and computability of the resulting abstract interpreter are then self-evident and easily proved.

The AAM approach has been applied to a wide variety of languages and applications, and given the success of the approach it's natural to wonder what is essential about its use of low-level machines. It is not at all clear whether a similar approach is possible with a higher-level formulation of the semantics, such as a compositional evaluation function defined recursively over the syntax of expressions.

This paper shows that the essence of the AAM approach can be applied to a high-level semantic basis. We show that compositional evaluators written in monadic style can express similar abstractions to that of AAM, and like AAM, the design remains systematic. Moreover, we show that the high-level semantics offers a number of benefits not available to the machine model.

There is a rich body of work concerning tools and techniques for *extensible* interpreters (Jaskelioff 2009; Kiselyov 2012; Liang et al. 1995), all of which applies to high-level semantics. By putting abstract interpretation for higher-order languages on a high-level semantic basis, we can bring these results to bear on the construction of extensible abstract interpreters.

## 3 A DEFINITIONAL INTERPRETER

We begin by constructing a definitional interpreter for a small but representative higher-order, functional language. The abstract syntax of the language is given below; it includes variables, numbers, binary operations on numbers, conditionals, recursive expressions, functions, and applications.

$$
\begin{array}{llll}
e \in exp & ::= & (\texttt{vbl}\ x) & [\textit{variable}] \\
& | & (\texttt{num}\ n) & [\textit{conditional}] \\
& | & (\texttt{if0}\ e\ e\ e) & [\textit{binary op}] \\
& | & (\texttt{app}\ e\ e) & [\textit{application}] \\
& | & (\texttt{rec}\ x\ e) & [\textit{rec binding}] \\
& | & (\texttt{lam}\ x\ e) & [\textit{function defn}] \\
x \in var & ::= & \texttt{x, y, ...} & [\textit{variable name}] \\
b \in bin & ::= & \texttt{+, -, ...} & [\textit{binary prim}]
\end{array}
$$

The interpreter for the language is defined in Figure 1. At first glance, it has many conventional aspects: it is compositionally defined by structural recursion on the syntax of expressions; it defines a call-by-value functional language, it represents function values as a closure data structure which pairs the lambda term with the evaluation environment; it is structured monadically and uses monad operations to interact with the environment and store; and it relies on a helper function $\delta$ to interpret primitive operations.

There are a few superficial aspects that deserve a quick note: environments $\rho$ are finite maps and the syntax $(\rho\ \texttt{x})$ denotes $\rho(x)$ while $(\rho\ \texttt{x}\ \texttt{a})$ denotes $\rho[x{\mapsto}a]$. Recursive expressions $(\texttt{rec f e})$ bind $\texttt{f}$ to the result of evaluating $\texttt{e}$ in the scope of $\texttt{e}$ itself; it is a run-time error if evaluating $\texttt{e}_0$ requires evaluating $\texttt{f}$. The $\texttt{do}$-notation is just shorthand for $\texttt{bind}$, as usual:

```
(define ((ev ev) e)
  (match e
    [(num n)        (return n)]
    [(vbl x)        (do ρ ← ask-env
                       (find (ρ x)))]
    [(if0 e₀ e₁ e₂) (do v  ← (ev e₀)
                       z? ← (zero? v)
                       (ev (if z? e₁ e₂)))]
    [(op2 o e₀ e₁)  (do v₀ ← (ev e₀)
                       v₁ ← (ev e₁)
                       (δ o v₀ v₁))]
    [(rec f e)      (do ρ  ← ask-env
                       a  ← (alloc f)
                       ρ′ = (ρ f a)
                       v  ← (local-env ρ′ (ev e))
                       (ext a v)
                       (return v))]
    [(lam x e₀)     (do ρ ← ask-env
                       (return (cons (lam x e₀) ρ)))]
    [(app e₀ e₁)    (do (cons (lam x e₂) ρ) ← (ev e₀)
                       v₁ ← (ev e₁)
                       a  ← (alloc x)
                       (ext a v₁)
                       (local-env (ρ x a) (ev e₂)))]]))
```

Figure 1: The Extensible Definitional Interpreter

$$(\text{do } x \leftarrow e \ . \ r) \equiv (\text{bind } e \ (\lambda \ (x) \ (\text{do } . \ r)))$$
$$(\text{do } e \ . \ r) \equiv (\text{bind } e \ (\lambda \ (\_) \ (\text{do } . \ r)))$$
$$(\text{do } x = e \ . \ r) \equiv (\text{let } ((x \ e)) \ (\text{do } . \ r))$$
$$(\text{do } b) \equiv b$$

Finally, there are two unconventional aspects worth noting.

First, the interpreter is written in an *open recursive style*; the evaluator does not call itself recursively, instead it takes as an argument a function `ev`—shadowing the name of the function `ev` being defined—and `ev` (the argument) is called instead of self-recursion. This is a standard encoding for recursive functions in a setting without recursive binding. It is up to an external function, such as the Y-combinator, to close the recursive loop. This open recursive form is crucial because it allows intercepting recursive calls to perform "deep" instrumentation of the interpreter.

Second, the code is clearly *incomplete*. There are a number of free variables, typeset as italics, which implement the following:

- The underlying monad of the interpreter: *return* and *bind*;
- An interpretation of primitives: $\delta$ and *zero?*;
- Environment operations: *ask-env* for retrieving the environment and *local-env* for installing an environment;
- Store operations: *ext* for updating the store, and *find* for dereferencing locations; and

- An operation for `alloc`ating new store locations.

Going forward, we make frequent use of definitions involving free variables, and we call such a collection of such definitions a *component*. We assume components can be named (in this case, we've named the component `ev@`, indicated by the box in the upper-right corner) and linked together to eliminate free variables.[1]

Next we examine a set of components which complete the definitional interpreter, shown in Figure 2. The first component `monad@` uses a macro `define-monad` which generates a set of bindings based on a monad transformer stack. We use a failure monad to model divide-by-zero errors, a state monad to model the store, and a reader monad to model the environment. The `define-monad` form generates bindings for `return`, `bind`, `ask-env`, `local-env`, `get-store` and `update-store`; their definitions are standard (Liang et al. 1995).

We also define `run` for monadic computations, starting in the empty environment and store:

```
(define (mrun m)
  (run-StateT ∅ (run-ReaderT ∅ m)))
```

While the `define-monad` form is hiding some details, this component could have equivalently been written out explicitly. For example, `return` and `bind` can be defined as:

```
(define (((return a) r) s) (cons a s))
(define (((bind ma f) r) s)
  (match ((ma r) s)
    [(cons a s′) (((f a) r) s′)]
    ['failure 'failure]))
```

So far our use of monad transformers is as a mere convenience, however the monad abstraction will become essential for easily deriving new analyses later on.

The $\delta@$ component defines the interpretation of primitives, which is given in terms of the underlying monad. The `alloc@` component provides `alloc`, which fetches the store and uses its size to return a fresh address, assuming the invariant $(\in \text{a } \sigma) \Leftrightarrow (< \text{a } (\text{size } \sigma))$. The `alloc` function takes a single argument, which is the name of the variable whose binding is being allocated. For the time being, it is ignored, but will become relevant when abstracting closures (section 3.3). The `store@` component defines `find` and `ext` for finding and extending values in the store.

The only remaining pieces of the puzzle are a fixed-point combinator and the main entry-point for the interpreter, which are straightforward to define:

```
(define ((fix f) x) ((f (fix f)) x))
(define (eval e) (mrun ((fix ev) e)))
```

By taking advantage of Racket's languages-as-libraries features (Tobin-Hochstadt et al. 2011), we construct REPLs for interacting with this interpreter. The following few evaluation examples demonstrate the interpreter working using a succinct concrete syntax. Here is a closure over the empty environment paired with the empty store and another over a non-empty environment, paired with a non-empty store:

```
> (λ (x) x)
'(((λ (x) x) . ()) . ())
```

Here is a closure over a non-empty environment and store:

```
> ((λ (x) (λ (y) x)) 4)
'(((λ (y) x) . ((x . 0))) . ((0 . 4)))
```

---

[1]We use Racket *units* (Flatt and Felleisen 1998) to model components in our implementation.

<div style="text-align: right">`monad@`</div>

```
(define-monad (ReaderT (FailT (StateT ID))))
```

<div style="text-align: right">`δ@`</div>

```
(define (δ o n₀ n₁)
  (match o
    ['+ (return (+ n₀ n₁))]
    ['- (return (- n₀ n₁))]
    ['* (return (* n₀ n₁))]
    ['/ (if (= 0 n₁) fail (return (/ n₀ n₁)))]))
(define (zero? v) (return (= 0 v)))
```

<div style="text-align: right">`store@`</div>

```
(define (find a)  (do σ ← get-store
                      (return (σ a))))
(define (ext a v) (update-store (λ (σ) (σ a v))))
```

<div style="text-align: right">`alloc@`</div>

```
(define (alloc x) (do σ ← get-store
                      (return (size σ))))
```

Figure 2: Components for Definitional Interpreters

<div style="text-align: right">`trace-monad@`</div>

```
(define-monad (ReaderT (FailT (StateT (WriterT List ID)))))
```

<div style="text-align: right">`ev-tell@`</div>

```
(define (((ev-tell ev₀) ev) e)
  (do ρ ← ask-env  σ ← get-store
      (tell (list e ρ σ))
      ((ev₀ ev) e)))
```

Figure 3: Trace Collecting Semantics

Primitive operations work as expected:

```
> (* (+ 3 4) 9)
'(63 . ())
```

Divide-by-zero errors result in failures:

```
> (/ 5 (- 3 3))
'(failure . ())
```

Because our monad stack places `FailT` above `StateT`, the answer includes the (empty) store at the point of the error. Had we changed `monad@` to use `(ReaderT (StateT (FailT ID)))` then failures would not include the store:

```
> (/ 5 (- 3 3))
'failure
```

```
                                                                              dead-monad@
(define-monad (ReaderT (StateT (StateT (FailT ID)))))

                                                                              ev-dead@

(define (((ev-dead ev₀) ev) e)
  (do θ ← get-dead
      (put-dead (set-remove θ e))
      ((ev₀ ev) e)))

                                                                              eval-dead@

(define ((eval-dead eval) e₀)
  (do (put-dead (subexps e₀))
      (eval e₀)))
```

Figure 4: Dead Code Collecting Semantics

At this point we've defined a simple definitional interpreter, although the extensible components involved—monadic operations and open recursion—will allow us to instantiate the same interpreter to achieve a wide range of useful abstract interpretations.

### 3.1 Collecting Variations

The formal development of abstract interpretation often starts from a so-called "non-standard collecting semantics." A common form of collecting semantics is a trace semantics, which collects streams of states the interpreter reaches. Figure 3 shows the monad stack for a tracing interpreter and a "mix-in" for the evaluator. The monad stack adds `WriterT List`, which provides a new operation `tell` for writing lists of items to the stream of reached states. The `ev-tell` function is a wrapper around an underlying `ev₀` unfixed evaluator, and interposes itself between each recursive call by `tell`ing the current state of the evaluator: the current expression, environment and store. The top-level evaluation function is then:

```
(define (eval e) (mrun ((fix (ev-tell ev)) e)))
```

Now when an expression is evaluated, we get an answer and a list of all states seen by the evaluator, in the order in which they were seen. For example:

```
> (* (+ 3 4) 9)
'((63 . ())
  ((* (+ 3 4) 9) () ())
  ((+ 3 4) () ())
  (3 () ())
  (4 () ())
  (9 () ())))
```

Were we to swap `List` with `Set` in the monad stack, we would obtain a *reachable* state semantics, another common form of collecting semantics, that loses the order and repetition of states.

As another collecting semantics variant, we show how to collect the *dead code* in a program. Here we use a monad stack that has an additional state component (with operations named `put-dead` and `get-dead`) which stores the set of dead expressions. Initially this will contain all subexpressions of the program. As the interpreter evaluates expressions it will remove them from the dead set.

Figure 4 defines the monad stack for the dead code collecting semantics and the `ev-dead@` component, another mix-in for an `ev₀` evaluator to remove the given subexpression before recurring. Since computing the dead code requires an outer wrapper that sets the initial set of dead code to be all of the subexpressions in the program, we define `eval-dead@` which consumes a *closed evaluator*, i.e. something of the form (`fix ev`). Putting these pieces together, the dead code collecting semantics is defined:

```
(define (eval e) (mrun ((eval-dead (fix (ev-dead ev))) e)))
```

Running a program with the dead code interpreter produces an answer and the set of expressions that were not evaluated during the running of a program:

```
> (if0 0 1 2)
(cons '(1 . ()) (set 2))
> (λ (x) x)
(cons '(((λ (x) x) . ()) . ()) (set 'x))
> (if0 (/ 1 0) 2 3)
(cons '(failure . ()) (set 3 2))
```

Our setup makes it easy not only to express the concrete interpreter, but also these useful forms of collecting semantics.

## 3.2 Abstracting Base Values

Our interpreter must become decidable before it can be considered an analysis, and the first step towards decidability is to abstract the base types of the language to something finite. We do this for our number base type by introducing a new *abstract* number, written `'N`, which represents the set of all numbers. Abstract numbers are introduced by an alternative interpretation of primitive operations, given in Figure 5, which simply produces `'N` in all cases.

Some care must be taken in the abstraction of `'/`. If the denominator is the abstract number `'N`, then it is possible the program could fail as a result of divide-by-zero, since `0` is contained in the interpretation of `'N`. Therefore there are *two* possible answers when the denominator is `'N`: `'N` and `'failure`. Both answers are `return`ed by introducing non-determinism `NondetT` into the monad stack. Adding non-determinism provides the `mplus` operation for combining multiple answers. Non-determinism is also used in `zero?`, which returns both true and false on `'N`.

By linking together $\delta\hat{}$@ and the monad stack with non-determinism, we obtain an evaluator that produces a set of results:

```
> (* (+ 3 4) 9)
'((N . ()))
> (/ 5 (+ 1 2))
'((failure . ()) (N . ()))
> (if0 (+ 1 0) 3 4)
'((4 . ()) (3 . ()))
```

If we link $\delta\hat{}$@ with the *tracing* monad stack plus non-determinism:

```
> (if0 (+ 1 0) 3 4)
(set
  '((4 . ()) (if0 (+ 1 0) 3 4) (+ 1 0) 1 0 4)
  '((3 . ()) (if0 (+ 1 0) 3 4) (+ 1 0) 1 0 3))
```

It is clear that the interpreter will only ever see a finite set of numbers (including `'N`) since the arguments to $\delta$ will only ever include numbers that appear in the program text or results of

`monad^@`

```
(define-monad (ReaderT (FailT (StateT (NondetT ID)))))
```

`δ^@`

```
(define (δ o n₀ n₁)
  (match* (o n₀ n₁)
    [('+ _ _)         (return 'N)]
    [('/ _ (? num?)) (if (= 0 n₁) fail (return 'N))]
    [('/ _ 'N)        (mplus fail (return 'N))] ...))
(define (zero? v)
  (match v
    ['N (mplus (return #t) (return #f))]
    [_  (return (= 0 v))]))
```

Figure 5: Abstracting Primitive Operations

`alloc^@`

```
(define (alloc x) (return x))
```

`store-nd@`

```
(define (find a)
  (do σ ← get-store
      (for/monad+ ([v (σ a)])
        (return v))))
(define (ext a v)
  (update-store (λ (σ) (σ a (if (∈ a σ) (set-add (σ a) v) (set v))))))
```

Figure 6: Abstracting Allocation: 0CFA

previous uses of $\delta$, which is just `'N`. However, it's definitely not true that the interpreter halts on all inputs. First, it's still possible to generate an infinite number of closures. Second, there's no way for the interpreter to detect when it sees a loop. To make a terminating abstract interpreter requires tackling both. We look next at abstracting closures.

### 3.3 Abstracting Closures

Closures consist of code—a lambda term—and an environment—a finite map from variables to addresses. Since the set of lambda terms and variables is bounded by the program text, it suffices to finitize closures by finitizing the set of addresses. Following the AAM approach, we do this by modifying the allocation function to produce elements drawn from a finite set. In order to retain soundness in the semantics, we modify the store to map addresses to *sets* of values, model store update as a join, and model dereference as a non-deterministic choice.

*Any* abstraction of the allocation function that produces a finite set will do (there's no way to make an unsound choice), but the choice of abstraction will determine the precision of the resulting analysis. A simple choice is to allocate variables using the variable's name as its address. This gives a monovariant, or 0CFA-like, abstraction.

Figure 6 shows the component `alloc^@` which implements monomorphic allocation, and the component `store-nd@` for implementing `find` and `ext` which interact with a store mapping to *sets* of values. The `for/monad+` form is a convenience for combining a set of computations with `mplus`, and is used so `find` returns *all* of the values in the store at a given address. The `ext` function joins whenever an address is already allocated, otherwise it maps the address to a singleton set. By linking these components with the same monad stack from before, we obtain an interpreter that loses precision whenever variables are bound to multiple values. For example, this program binds `x` to both `1` and `2` and produces both answers when run:

```
> (let ((f (λ (x) x)))
    (f 1)
    (f 2))
'(1 2)
```

Our abstract interpreter now has a truly finite domain; the next step is to detect loops in the state-space to achieve termination.

## 4   CACHING AND FINDING FIXED-POINTS

At this point, the interpreter obtained by linking together `monad^@`, $\delta$`^@`, `alloc^@` and `store-nd@` components will only ever visit a finite number of configurations for a given program. A configuration ($\varsigma$) consists of an expression (e), environment ($\rho$) and store ($\sigma$). This configuration is finite because: expressions are finite in the given program; environments are maps from variables (again, finite in the program) to addresses; the addresses are finite thanks to `alloc^`; the store maps addresses to sets of values; base values are abstracted to a finite set by $\delta$`^`; and closures consist of an expression and environment, which are both finite.

Although the interpreter will only ever see a finite set of inputs, it *doesn't know it*. A simple loop will cause the interpreter to diverge:

```
> ((rec f (λ (x) (f x))) 0)
with-limit: out of time
```

To solve this problem, we introduce a *cache* (`$in`) as input to the algorithm, which maps from configurations ($\varsigma$) to sets of value-and-store pairs ($v \times \sigma$). When a configuration is reached for the second time, rather than re-evaluating the expression and entering an infinite loop, the result is looked up from `$in`, which acts as an oracle. It is important that the cache is used in a *productive* way: it is only safe to use `in` as an oracle so long as some progress has been made first.

The results of evaluation are then stored in an output cache (`$out`), which after the end of evaluation is "more defined" than the input cache (`$in`), again following a co-inductive argument. The least fixed-point $\$^+$ of an evaluator which transforms an oracle `$in` and outputs a more defined oracle $racket[out] is then a sound approximation of the program, because it over-approximates all finite unrollings of the unfixed evaluator.

The co-inductive caching algorithm is shown in Figure 7, along with the monad transformer stack `monad-cache@` which has two new components: `ReaderT` for the input cache `$in`, and `StateT+` for the output cache `$out`. We use a `StateT+` instead of `WriterT` monad transformer in the output cache so it can double as tracking the set of seen states. The `+` in `StateT+` signifies that caches for multiple non-deterministic branches will be merged automatically, producing a set of results and a single cache, rather than a set of results paired with individual caches.

In the algorithm, when a configuration $\varsigma$ is first encountered, we place an entry in the output cache mapping $\varsigma$ to (`$in` $\varsigma$), which is the "oracle" result. Also, whenever we finish computing the result $v \times \sigma\prime$ of evaluating a configuration $\varsigma$, we place an entry in the output cache mapping $\varsigma$

monad-cache@

```
(define-monad
  (ReaderT (FailT (StateT (NondetT (ReaderT (StateT+ ID)))))))
```

ev-cache@

```
(define (((ev-cache ev₀) ev) e)
  (do ρ ← ask-env
      σ ← get-store
      ς = (list e ρ σ)
      $out ← get-cache-out
      (if (∈ ς $out)
          (for/monad+ ([v×σ ($out ς)])
            (do (put-store (cdr v×σ))
                (return (car v×σ))))
          (do $in ← ask-cache-in
              v×σ₀ = (if (∈ ς $in) ($in ς) ∅)
              (put-cache-out ($out ς v×σ₀))
              v ← ((ev₀ ev) e)
              σ′ ← get-store
              v×σ′ = (cons v σ′)
              (update-cache-out
               (λ ($out) ($out ς (set-add ($out ς) v×σ′))))
              (return v)))))
```

Figure 7: Co-inductive Caching Algorithm

to v×σ′. Finally, whenever we reach a configuration ς for which a mapping in the output cache exists, we use it immediately, returning each result using the for/monad+ iterator. Therefore, every "cache hit" on $out is in one of two possible states: 1) we have already seen the configuration, and the result is the oracle result, as desired; or 2) we have already computed the "improved" result (w.r.t. the oracle), and need not recompute it.

To compute the least fixed-point $$^+$ for the evaluator ev-cache we perform a standard Kleene fixed-point iteration starting from the empty map, the bottom element for the cache, as shown in Figure 8.

The algorithm iterates the caching evaluator eval on the given program e from the initial environment and store. The monadic least fixed-point finder mlfp evaluates the program with an empty cache and the improved oracle $. The caching eval populates the cache as it evaluates, including results from the oracle when configurations are known from prior iterations, and returns the cache to be used as the improved oracle during the next iteration. After finding the least fixed-point of the oracle, the final values and store for the initial configuration ς are extracted and returned.

Termination of the least fixed-point is justified by the monotonicity of the evaluator (it always returns an "improved" oracle), and the finite domain of the cache, which maps abstract configurations to pairs of values and stores, all of which are finite.

With these pieces in place we construct a complete interpreter:

```
(define (eval e) (mrun ((fix-cache (fix (ev-cache ev))) e)))
```

```
                                                                    fix-cache@

(define ((fix-cache eval) e)
  (do ρ ← ask-env   σ ← get-store
      ς = (list e ρ σ)
      $⁺ ← (mlfp (λ ($) (do (put-cache-out ∅)
                            (put-store σ)
                            (local-cache-in $ (eval e))
                            get-cache-out)))
      (for/monad+ ([v×σ ($⁺ ς)])
        (do (put-store (cdr v×σ))
            (return (car v×σ))))))))
(define (mlfp f)
  (let loop ([x ∅])
    (do x′ ← (f x)
        (if (equal? x′ x) (return x) (loop x′)))))
```

Figure 8: Finding Fixed-Points in the Cache

When linked with $δ^{\wedge}$ and `alloc^`, this abstract interpreter is sound and computable, as demonstrated on the following examples:

```
> ((rec f (λ (x) (f x)))
   0)
'()
> ((rec f (λ (n) (if0 n 1 (* n (f (- n 1))))))
   5)
'(N)
> ((rec f (λ (x) (if0 x 0 (if0 (f (- x 1)) 2 3))))
   (+ 1 0))
'(2 0 3)
```

**Formal soundness and termination**

In this pearl, we have focused on the code and its intuitions rather than rigorously establishing the usual formal properties of our abstract interpreter, but this is just a matter of presentation: the interpreter is indeed proven sound and computable. We have formalized this co-inductive caching algorithm in the supplemental material accompanying this paper, where we prove both that it always terminates, and that it computes a sound over-approximation of concrete evaluation. Here, we give a short summary of our metatheory approach.

In formalising the soundness of this caching algorithm, we extend a standard big-step evaluation semantics into a *big-step reachability* semantics, which characterizes all intermediate configurations which are seen between the evaluation of a single expression and its eventual result. These two notions—*evaluation* which relates expressions to fully evaluated results, and *reachability* which characterizes intermediate configuration states—remain distinct throughout the formalism.

After specifying evaluation and reachability for concrete evaluation, we develop a *collecting* semantics which gives a precise specification for any abstract interpreter, and an *abstract* semantics which partially specifies a sound, over-approximating algorithm w.r.t. the collecting semantics.

The final step is to compute an oracle for the *abstract evaluation relation*, which maps individual configurations to abstractions of the values they evaluate to. To construct this cache, we *mutually* compute the least-fixed point of both the evaluation and reachability relations: based on what is evaluated, discover new things which are reachable, and based on what is reachable, discover new results of evaluation. The caching algorithm developed in this section is a slightly more efficient strategy for solving the mutual fixed-point, by taking a deep exploration of the reachability relation (up-to seeing the same configuration twice) rather than applying just a single rule of inference.

## 5 PUSHDOWN *À LA* REYNOLDS

By combining the finite abstraction of base values and closures with the termination-guaranteeing cache-based fixed-point algorithm, we have obtained a terminating abstract interpreter. But what kind of abstract interpretation did we get? We have followed the basic recipe of AAM, but adapted to a compositional evaluator instead of an abstract machine. However, we did manage to skip over one of the key steps in the AAM method: we never store-allocated continuations. *In fact, there are no continuations at all!*

A traditional abstract machine formulation of the semantics would model the object-level stack explicitly as an inductively defined data structure. Because stacks may be arbitrarily large, they must be finitized like base values and closures, and like closures, the AAM trick is to thread them through the store, which itself must become finite. But in the definitional interpreter approach, the story of this paper, the model of the stack is implicit and simply inherited from the meta-language.

But here is the remarkable thing: since the stack is inherited from the meta-language, the abstract interpreter inherits the "call-return matching" of the meta-language, which is to say there is no loss of precision of in the analysis of the control stack. This is a property that usually comes at considerable effort and engineering in the formulations of higher-order flow analysis that model the stack explicitly. So-called higher-order "pushdown" analysis has been the subject of multiple publications and two dissertations (Earl 2014; Earl et al. 2010; Earl et al. 2012; Gilray et al. 2016; Johnson et al. 2014; Johnson and Van Horn 2014; Van Horn and Might 2012; Vardoulakis 2012; Vardoulakis and Shivers 2011). Yet when formulated in the definitional interpreter style, the pushdown property requires no mechanics and is simply inherited from the meta-language.

Reynolds, in his celebrated paper *Definitional Interpreters for Higher-order Programming Languages* (Reynolds 1972), first observed that when the semantics of a programming language is presented as a definitional interpreter, the defined language could inherit semantic properties of the defining metalanguage. We have now shown this observation can be extended to *abstract* interpretation as well, namely in the important case of the pushdown property.

In the remainder of this paper, we explore a few natural extensions and variations on the basic pushdown abstract interpreter we have established up to this point.

## 6 WIDENING THE STORE

In this section, we show how to recover the well-known technique of store-widening and in doing so demonstrate the ease with which we can construct existing abstraction design choices.

The abstract interpreter we've constructed so far uses a store-per-program-state abstraction, which is precise but prohibitively expensive. A common technique to combat this cost is to use a global "widened" store (Might 2007a; Shivers 1991), which over-approximates each individual store in the current set-up. This change is achieved easily in the monadic setup by re-ordering the monad stack, a technique due to Darais et al. (2015). Whereas before we had `monad-cache@` we instead swap the order of `StateT` for the store and `NondetT`:

```
(ReaderT (FailT (NondetT (StateT+ (ReaderT (StateT+ ID))))))
```

<div align="right">

`precise-`$\delta$`@`
</div>

```
(define (δ o n₀ n₁)
  (match* (o n₀ n₁)
    [('+ (? num?) (? num?)) (return (+ n₀ n₁))]
    [('+ _        _)        (return 'N)] ...))
(define (zero? v)
  (match v
    ['N (mplus (return #t) (return #f))]
    [_  (return (= 0 v))]))
```

<div align="right">

`store-crush@`
</div>

```
(define (find a)
  (do σ ← get-store
      (for/monad+ ([v (σ a)])
        (return v))))
(define (crush v vs)
  (if (closure? v)
      (set-add vs v)
      (set-add (set-filter closure? vs) 'N)))
(define (ext a v)
  (update-store (λ (σ) (if (∈ a σ)
                           (σ a (crush v (σ a)))
                           (σ a (set v))))))
```

Figure 9: An Alternative Abstraction for Precise Primitives

we get a store-widened variant of the abstract interpreter. Because `StateT` for the store appears underneath nondeterminism, it will be automatically widened. We write `StateT+` to signify that the cell of state supports such widening.

## 7  AN ALTERNATIVE ABSTRACTION

In this section, we demonstrate how easy it is to experiment with alternative abstraction strategies by swapping out components. In particular we look at an alternative abstraction of primitive operations and store joins that results in an abstraction that—to the best of our knowledge—has not been explored in the literature. This example shows the potential for rapidly prototyping novel abstractions using our approach.

Figure 9 defines two new components: `precise-`$\delta$`@` and `store-crush@`. The first is an alternative interpretation for primitive operations that is *precision preserving*. Unlike $\delta$`^@`, it does not introduce abstraction, it merely propagates it. When two concrete numbers are added together, the result will be a concrete number, but if either number is abstract then the result is abstract.

This interpretation of primitive operations clearly doesn't impose a finite abstraction on its own, because the state space for concrete numbers is infinite. If `precise-`$\delta$`@` is linked with the `store-nd@` implementation of the store, termination is therefore not guaranteed.

The `store-crush@` operations are designed to work with `precise-`$\delta$`@` by performing *widening* when joining multiple concrete values into the store. This abstraction offers a high-level of precision; for example, constant arithmetic expressions are computed with full precision:

```
> (* (+ 3 4) 9)
'(63)
```

Even linear binding and arithmetic preserves precision:

```
> ((λ (x) (* x x)) 5)
'(25)
```

Only when the approximation of binding structure comes in to contact with base values that we see a loss in precision:

```
> (let ((f (λ (x) x)))
    (* (f 5) (f 5)))
'(N)
```

This combination of `precise-δ@` and `store-crush@` allows termination for most programs, but still not all. In the following example, `id` is eventually applied to a widened argument `'N`, which makes both conditional branches reachable. The function returns `0` in the base case, which is propagated to the recursive call and added to `1`, which yields the concrete answer `1`. This results in a cycle where the intermediate sum returns `2`, `3`, `4` when applied to `1`, `2`, `3`, etc.

```
> ((rec id (λ (n) (if0 n 0 (+ 1 (id (- n 1))))))
    3)
with-limit: out of time
```

To ensure termination for all programs, we assume all references to primitive operations are $\eta$-expanded, so that store-allocations also take place at primitive applications, ensuring widening at repeated bindings. In fact, all programs terminate when using `precise-δ@`, `store-crush@` and $\eta$-expanded primitives, which means we have a achieved a computable and uniquely precise abstract interpreter.

Here we see one of the strengths of the extensible, definitional approach to abstract interpreters. The combination of added precision and widening is encoded quite naturally. In contrast, it's hard to imagine how such a combination could be formulated as, say, a constraint-based flow analysis.

## 8  SYMBOLIC EXECUTION

In this section, we carry out another—this time more involved—example that shows how to instantiate our definitional abstract interpreter to obtain a symbolic execution engine that performs sound program verification. This serves to demonstrate the range of the approach, capturing forms of analysis typically considered fairly dissimilar.

First, we describe the monad stack and metafunctions that implement a symbolic executor (King 1976), then we show how abstractions discussed in previous sections can be applied to enforce termination, turning a traditional symbolic execution into a path-sensitive verification engine.

To support symbolic execution, the syntax of the language is extended to include symbolic numbers:

$$
\begin{array}{llll}
e \in exp & ::= \ldots \mid (\text{sym } x) & [\textit{symbolic number}] \\
\varepsilon \in pexp & ::= e \mid \neg e & [\textit{path expression}] \\
\phi \in pcon & ::= P(pexp) & [\textit{path condition}]
\end{array}
$$

Figure 11 shows the units needed to turn the existing interpreter into a symbolic executor. Primitives such as `'/` now also take as input and return symbolic values. As standard, symbolic execution employs a path-condition accumulating assumptions made at each branch, allowing the elimination of provably infeasible paths and construction of test cases. We represent the path-condition $\phi$ as a set of symbolic values or their negations. If `e` is in $\phi$, `e` is assumed to evaluate to `0`; if $\neg$ `e`

`symbolic-monad@`

```
(define-monad (ReaderT (FailT (StateT (StateT (NondetT ID))))))
```

`ev-symbolic@`

```
(define (((ev-symbolic ev₀) ev) e)
  (match e
    [(sym x) (return x)]
    [_       ((ev₀ ev) e)]))
```

`δ-symbolic@`

```
(define (δ o n₀ n₁)
  (match* (o n₀ n₁)
    [('/ n₀ n₁) (do z? ← (zero? n₁)
                    (cond [z? fail]
                          [(and (num? n₀) (num? n₁)) (return (/ n₀ n₁))]
                          [else (return `(/ ,n₀ ,n₁))]))] ...))
(define (zero? v)
  (do φ ← get-path-cond
      (match v
        [(? num? n)               (return (= 0 n))]
        [v #:when (∈ v φ)         (return #t)]
        [v #:when (∈ `(¬ ,v) φ)   (return #f)]
        [v (mplus (do (refine v)        (return #t))
                  (do (refine `(¬ ,v)) (return #f)))])))
```

Figure 10: Symbolic Execution Variant

`δ^-symbolic@`

```
(define (δ o n₀ n₁)
  (match* (o n₀ n₁)
    [('/ n₀ n₁) (do z? ← (zero? n₁)
                    (cond [z? fail]
                          [(member 'N (list n₀ n₁)) (return 'N)]
                          ...))]
    ...))
(define (zero? v)
  (do φ ← get-path-cond
      (match v ['N (mplus (return #t) (return #f))] ...)))
```

Figure 11: Symbolic Execution with Abstract Numbers

is in $\phi$, e is assumed to evaluate to non-$0$. This set is another state component provided by `StateT` in the monad transformer stack. Monadic operations `get-path-cond` and `refine` reference and update the path-condition. The metafunction `zero?` works similarly to the concrete counterpart, but also uses the path-condition to prove that some symbolic numbers are definitely $0$ or non-$0$.

In case of uncertainty, `zero?` returns both answers instead of refining the path-condition with the assumption made.

In the following example, the symbolic executor recognizes that result `3` and division-by-0 error are not feasible:

```
> (if0 'x (if0 'x 2 3) (/ 5 'x))
(set
 (cons 2 (set 'x))
 (cons '(/ 5 x) (set '(¬ x))))
```

A scaled up symbolic executor could implement `zero?` by calling out to an SMT solver for more interesting reasoning about arithmetic, or extend the language with symbolic functions and blame semantics for sound higher-order symbolic execution, essentially recreating a pushdown variant of Nguyễn et al. (Nguyễn et al. 2014; Nguyễn and Van Horn 2015; Tobin-Hochstadt and Van Horn 2012).

Traditional symbolic executors aim to find bugs and do not provide a termination guarantee. However, if we apply the finite abstractions and caching of the previous sections (section 3.2 and section 4) to this symbolic executor, we turn it into a sound, path-sensitive verification engine.

There is one wrinkle, which is that operations on symbolic values introduce a new source of unboundness in the state-space, because the space of symbolic values is not finite. A simple strategy to ensure termination is to widen a symbolic value to the abstract number `'N` when it shares an address with a different number, similarly to the precision-preserving abstraction from section 7. Figure 11 shows extension to $\delta$ and `zero?` in the presence of `'N`. The different treatments of `'N` and symbolic values clarifies that abstract values are not symbolic values: the former stands for a set of multiple values, whereas the latter stands for an single unknown value. Tests on abstract number `'N` do not strengthen the path-condition; it is unsound to accumulate any assumption about `'N`.

## 9  GARBAGE COLLECTION

As a denouement to our series of examples, we show how to incorporate garbage collection into our definitional abstract interpreter.

This example, like store-widening, is the re-creation of a well-known technique: abstract garbage collection (Might and Shivers 2006b) mimics the process of reclaiming unreachable heap addresses as done in garbage-collecting concrete interpreters. While garbage collection in the concrete can largely be considered an implementation detail that doesn't effect the results of computation (modulo pragmatic issues of memory consumption), in the abstract semantics, it can significantly improve the precision of analysis results. This is because store locations mediate joins, and therefore imprecision, in the abstract semantics. If an address can be collected, it avoids a later join that would otherwise be encountered without garbage collection.

In the finite-state-machine model, abstract garbage collection is fairly straightforward and closely follows concrete formulations (Might and Shivers 2006b; Van Horn and Might 2010). However, incorporating both pushdown control flow and abstract garbage collection has proved rather involved and required new techniques (Earl et al. 2012; Johnson et al. 2014). The key difficulty for pushdown machine models, which essentially use abstract models that are pushdown automata, is that the usual approach to garbage collection is to crawl the call stack to compute the root set of reachable addresses (Morrisett et al. 1995). Traversing the stack, however, is not something that can be expressed by a pushdown automata. This difficulty is somewhat exacerbated by the definitional interpreter approach in combination with a metalanguage (Racket) that doesn't reify a stack to traverse! Nevertheless, as we demonstrate, this challenge can be overcome to obtain a pushdown, garbage-collecting abstract interpreter. Doing so shows that the definitional abstract

```
                                                          monad-pdcfa-gc@
(define-monad
   (ReaderT (ReaderT (FailT (StateT (NondetT (ReaderT (StateT+ ID)))))))))

                                                          mrun-pdcfa-gc@
(define (mrun m)
   (run-StateT+ ∅ (run-ReaderT ∅    ; out-$0, in-$0
   (run-StateT  ∅ (run-ReaderT ∅    ; σ0, ρ0
   (run-ReaderT (set) m))))))       ; ψ0
```

Figure 12: Monad Instance with Root Address Set

interpreter approach also scales to handle so-called *introspective* pushdown analysis that require some level of introspection on the stack (Earl et al. 2012; Johnson et al. 2014).

Solving the abstract garbage collection problem boils down to answering the following question: how can we track root addresses that are live on the call stack when the call stack is implicitly defined by the metalanguage? The answer is fairly simple: we extend the monad with a set of root addresses. When evaluating compound expressions, we calculate the appropriate root sets for the context. In essence, we render explicit only the addresses of the calling context, while still relying on the metalanguage to implicitly take care of the rest as before.

Figure 12 defines the appropriate monad instance. All that has changed is there is an added reader component, which will be used to model the context's current root set. The use of this added component necessitates a change to the caching and fixed-point calculation, namely we must include the root sets as part of the configuration. Compared with the `ev-cache@` component of section 4, we make a simple adjustment to the first few lines to cache the root set along with the rest of the configuration:

```
(define (((ev-cache ev0) ev) e)
   (do ρ ← ask-env  σ ← get-store  ψ ← ask-roots
       ς = (list e ρ σ ψ)
       ...))
```

Similarly, for `fix-cache@`:

```
(define ((fix-cache eval) e)
   (do ρ ← ask-env  σ ← get-store  ψ ← ask-roots
       ς = (list e ρ σ ψ)
       ...))
```

We can now write a `ev-collect@` component that collects garbage: it asks for the current roots in the context, evaluates an expression to a value, then updates the store after collecting all addresses not reachable from the roots of the context and value:

```
(define (((ev-collect ev0) ev) e)
   (do ψ ← ask-roots
       v ← ((ev0 ev) e)
       (update-store (gc (set-union ψ (roots-v v))))
       (return v)))
```

Here, `gc` and `roots-v` are (omitted) helper functions that perform garbage collection and calculate the set of root addresses in a value, respectively.

ev-roots@

```
(define (((ev-roots ev₀) ev) e)
  (match e
    [(if0 e₀ e₁ e₂) (do ρ  ← ask-env
                        ψ′ = (set-union (roots e₁ ρ) (roots e₂ ρ))
                        v  ← (extra-roots ψ′ (ev e₀))
                        b  ← (truish? v)
                        (ev (if b e₁ e₂)))]
    [(op2 o e₀ e₁)  (do ρ ← ask-env
                        v₀ ← (extra-roots (roots e₁ ρ) (ev e₀))
                        v₁ ← (extra-roots (roots-v v₀) (ev e₁))
                        (δ o v₀ v₁))]
    [(app e₀ e₁)    (do ρ  ← ask-env
                        v₀ ← (extra-roots (roots e₁ ρ) (ev e₀))
                        v₁ ← (extra-roots (roots-v v₀) (ev e₁))
                        (cons (lam x e₂) ρ′) = v₀
                        a  ← (alloc x)
                        (ext a v₁)
                        (local-env (ρ′ x a) (ev e₂)))]
    [_ ((ev₀ ev) e)]))
```

Figure 13: Address Collection and Propagation

All that remains is to define a component that propagates root sets appropriately from compound expressions to their constituents. Figure 13 gives the `ev-roots@` component, which does exactly this. Finally, the pieces are stitched together with the following to obtain a pushdown, garbage-collecting definitional abstract interpreter:

```
(define (eval e)
  (mrun ((fix-cache (fix (ev-cache (ev-collect (ev-roots ev))))) e)))
```

To observe the added precision due to GC, consider the following example, run using the (non-garbage-collecting) pushdown abstract interpreter of section 5:

```
> (let ((f (λ (x) x)))
    (f 1)
    (f 2))
'(1 2)
```

This example binds `f` to an identity function and applies `f` to two arguments, 1 and 2. Since the first binding of `x` to 1 is still in the store when the second binding to 2 happens, the results are joined. This causes the second application of `f` to produce *both* 1 and 2. If instead the garbage-collecting variant is used, there is a collection between the two calls to `f`, which is after the first binding of `x` but before the second. At this moment, `x` is unreachable and collected. When `f` is applied again, `x` gets bound in a fresh location to *just* 2 and the overall result reflects this more precise fact:

```
> (let ((f (λ (x) x)))
    (f 1)
    (f 2))
'(2)
```

## 10 RELATED WORK

This work draws upon and re-presents many ideas from the literature on abstract interpretation for higher-order languages (Midtgaard 2012). In particular, it closely follows the abstracting abstract machines (Van Horn and Might 2010, 2012) approach to deriving abstract interpreters from a small-step machine. The key difference here is that we operate in the setting of a monadic definitional interpreter instead of an abstract machine. In moving to this new setting we developed a novel caching mechanism and fixed-point algorithm, but otherwise followed the same recipe. Remarkably, in the setting of definitional interpreters, the pushdown property for the analysis is simply inherited from the meta-language rather than requiring explicit instrumentation to the abstract interpreter.

Compositionally defined abstract interpretation functions for higher-order languages were first explored by Jones and Nielson (1995), which introduces the technique of interpreting a higher-order object language directly as terms in a meta-language to perform abstract interpretation. While their work lays the foundations for this idea, it does not consider abstractions for fixed-points in the domain, so although their abstract interpreters are sound, they are not in general computable. They propose a naïve solution of truncating the interpretation of syntactic fixed-points to some finite depth, but this solution isn't general and doesn't account for non-syntactic occurrences of bottom in the concrete domain (*e.g. via* Y combinators). Our work develops such an abstraction for concrete denotational fixed-points using a fixed-point caching algorithm, resulting in general, computable abstractions for arbitrary definitional interpreters.

Perhaps the mostly closely related work is an unpublished graduate-level tutorial by Friedman and Medhekar (2003), which presents an "abstracted" interpreter parameterized over the interpretation of (abstract) semantic domains and written in an open recursive style, which is then closed using a non-standard, caching fixpoint operator to ensure termination. The interpreter closely resembles the definitional interpreter of section 3 and the caching fixed-point operator is similar to that of section 4. In contrast, their interpreter is not written in a monadic style for exensibility and uses mutation to model the cache and communicate updates during abstract interpretation. The latter makes it difficult to reason about the algorithm. The notes present no argument for soundness or termination, although both are implied. Notably, the interpreter seems to include no abstraction of closures and relies instead on the cache to prevent non-termination whenever a previously encountered closure is seen. Unfortunately, the strategy appears to be unsound—computations invovling Church-numeral style iteration produce "⊥," but don't diverge when run concretely. Nevertheless, the tutorial is remarkably close in spirit to the present work.

The use of monads and monad transformers to make extensible (concrete) interpreters is a well-known idea (Liang et al. 1995; Moggi 1989; Steele 1994), which we have extended to work for compositional abstract interpreters. The use of monads and monad transformers in machine based-formulations of abstract interpreters has previously been explored by Sergey et al. (2013) and Darais et al. (2015), respectively, and inspired our own adoption of these ideas. Darais has also shown that certain monad transformers are also *Galois transformers*, i.e. they compose to form monads that transport Galois connections. This idea may pave a path forward for obtaining both compositional code *and proofs* for abstract interpreters in the style presented here.

The caching mechanism used to ensure termination in our abstract interpreter is similar to that used by Johnson and Van Horn (2014). They use a local- and meta-memoization table in a machine-based interpreter to ensure termination for a pushdown abstract interpreter. This mechanism is in turn reminiscent of the use of memoization in an interpreter for two-way non-deterministic pushdown automata by Glück (2013).

Caching recursive, non-deterministic functions is a well-studied problem in the functional logic programming community under the rubric of "tabling" (Bol and Degerstedt 1993; Chen and Warren 1996; Swift and Warren 2012; Tamaki and Sato 1986), and has been usefully applied to program verification and analysis (Dawson et al. 1996; Janssens and Sagonas 1998). Unlike these systems, our approach uses a shallow embedding of cached non-determinism that can be applied in general-purpose functional languages. Monad transformers that enable shallow embedding of cached non-determinism are of continued interest since Hinze's *Deriving Backtracking Monad Transformers* (Fischer et al. 2011; Hinze 2000; Kiselyov et al. 2005), and recent work (Ploeg and Kiselyov 2014; Vandenbroucke et al. 2016) points to potential optimizations and specializations that can be applied to our relatively naive iteration strategy.

Vardoulakis, who was the first to develop the idea of a pushdown abstraction for higher-order flow analysis (Vardoulakis and Shivers 2011), formalized CFA2 using a CPS model, which is similar in spirit to a machine-based model. However, in his dissertation (Vardoulakis 2012) he sketches an alternative presentation dubbed "Big CFA2" which is a big-step operational semantics for doing pushdown analysis quite similar in spirit to the approach presented here. One key difference is that Big CFA2 fixes a particular coarse abstraction of base values and closures—for example, both branches of a conditional are always evaluated. Consequently, it only uses a single iteration of the abstract evaluation function, and avoids the need for the cache-based fixed-point of section 4. We believe Big CFA2 as stated is sound, however if the underlying abstractions were tightened, it may then require a more involved fixed-point finding algorithm like the one we developed.

Our formulation of a pushdown abstract interpreter computes an abstraction similar to the many existing variants of pushdown flow analysis (Earl et al. 2010; Earl et al. 2012; Gilray et al. 2016; Johnson et al. 2014; Johnson and Van Horn 2014; Van Horn and Might 2012; Vardoulakis 2012; Vardoulakis and Shivers 2011).

The mixing of symbolic execution and abstract interpretation is similar in spirit to the *logic flow analysis* of Might (Might 2007b), albeit in a pushdown setting and with a stronger notion of negation; generally, our presentation resembles traditional formulations of symbolic execution more closely (King 1976). Our approach to symbolic execution only handles the first-order case of symbolic values, as is common. However, Nguyễn's work on higher-order symbolic execution (Nguyễn and Van Horn 2015) demonstrates how to scale to behavioral symbolic values. In principle, it should be possible to handle this case in our approach by adapting Nguyễn's method to a formulation in a compositional evaluator, but this remains to be carried out.

Now that we have abstract interpreters formulated with a basis in abstract machines and with a basis in monadic interpreters, an obvious question is can we obtain a correspondence between them similar to the functional correspondence between their concrete counterparts (Ager et al. 2005). An interesting direction for future work is to try to apply the usual tools of defunctionalization, CPS, and refocusing to see if we can interderive these abstract semantic artifacts.

## 11 CONCLUSIONS

We have shown that the AAM methodology can be adapted to definitional interpreters written in monadic style. Doing so captures a wide variety of semantics, such as the usual concrete semantics, collecting semantics, and various abstract interpretations. Beyond recreating existing techniques from the literature such as store-widening and abstract garbage collection, we can also design novel abstractions and capture disparate forms of program analysis such as symbolic execution. Further, our approach enables the novel combination of these techniques.

To our surprise, the definitional abstract interpreter we obtained implements a form of pushdown control flow abstraction in which calls and returns are always properly matched in the abstract semantics. True to the definitional style of Reynolds, the evaluator involves no explicit mechanics to achieve this property; it is simply inherited from the metalanguage.

We believe this formulation of abstract interpretation offers a promising new foundation towards re-usable components for the static analysis and verification of higher-order programs. Moreover, we believe the definitional abstract interpreter approach to be a fruitful new perspective on an old topic. We are left wondering: what else can be profitably inherited from the metalanguage of an abstract interpreter?

## ACKNOWLEDGMENTS

The seeds of this work were planted while at the Northeastern University Programming Research Laboratory and the ideas benefited greatly from lively discussions within the PRL. In particular, we thank Sam Tobin-Hochstadt and Dionna Glaze for several fruitful conversations. We also thank the people of the Laboratory for Programming Languages at the University of Maryland for help developing these ideas more recently. Finally, we are grateful for the constructive feedback from the anonymous reviewers of ICFP 2016 and 2017. We thank Reviewer B of ICFP 2017 in particular for the reference to the Friedman and Medhekar (2003) tutorial, *Using an Abstracted Interpreter to Understand Abstract Interpretation*, an (unknown to us) precursor to the current work that anticipated our use of interpreters written using open recursion to explain abstract intepretation. This work was supported in part by NSF grants #1618756 and #1518765.